\documentclass{aa} 
\usepackage{graphicx, txfonts, natbib}
\bibpunct{(}{)}{;}{a}{}{,}

\newcommand{\hneut}{\mbox{$\mathrm{H}$}}
\newcommand{\hmol}{\mbox{$\mathrm{H_2}$ }}
\newcommand{\hmin}{\mbox{$\mathrm{H^-}$ }}
\newcommand{\dcr}{\mbox{$D_\mathrm{cr}$ }}
\newcommand{\dmw}{\mbox{$D_\mathrm{MW}$ }}
\newcommand{\yhmol}{\mbox{$y_\mathrm{H_2}$ }}
\newcommand{\yhmin}{\mbox{$y_\mathrm{H^-}$ }}
\newcommand{\dsfr}{\mbox{ $\mathrm{[M_{\sun}/yr/kpc^2]}$}}
\newcommand{\dkel}{\mbox{ $\mathrm{K}$}}
\newcommand{\jmass}{\mbox{$M_\mathrm{J}$}}
\newcommand{\jrad}{\mbox{$r_\mathrm{J}$}}
\newcommand{\tcool}{\mbox{$t_\mathrm{cool}$}}
\newcommand{\sfr}{\mbox{$\mathrm{SFR}$}}
\newcommand{\sfe}{\mbox{$\mathrm{SFE}$}}
\newcommand{\teff}{\mbox{$T_\mathrm{eff}$}}

\begin{document}

\title{The self regulating star formation of gas rich dwarf galaxies in
quiescent phase}

\author{Masakazu A.R. Kobayashi 
  \and Hideyuki Kamaya}

\institute{Department of Astronomy, School of Science, Kyoto University,
Sakyo-ku, Kyoto 606-8502, JAPAN}


\offprints{M. A.R. Kobayashi,\\
 \email{kobayasi@kusastro.kyoto-u.ac.jp}}

\date{Received 2 July 2004 / Accepted 30 August 2004}

\abstract{
 The expected episodic or intermittent star formation histories (SFHs)
 of gas rich dwarf irregular galaxies (dIrrs) are the longstanding
 puzzles to understand their whole evolutional history. Solving this
 puzzle, we should grasp what physical mechanism causes the quiescent
 phase of star formation under the very gas rich condition after the
 first starburst phase. We consider that this quiescent phase is kept by
 lack of $\mathrm{H_2}$, which can be important coolant to generate the
 next generation of stars in the low-metal environment like
 dIrrs. Furthermore, in dIrrs, \hmol formation through gas-phase
 reactions may dominate the one on dust-grain surfaces because their
 interstellar medium (ISM) are very plentiful and the typical
 dust-to-gas ratio of dIrrs ($D_\mathrm{dIrrs} = 1.31\times
 10^{-2}D_\mathrm{MW}$, where \dmw is its value for the local ISM) is on
 the same order with a critical value $D_\mathrm{cr} \sim
 10^{-2}D_\mathrm{MW}$. We show that the lack of \hmol is mainly led by
 \hmin destruction when gas-phase \hmol formation dominates since \hmin
 is important intermediary of gas-phase \hmol formation. \hmin is
 destroyed by the radiation from all stars born in the previous
 starburst phase because \hmin destroying infrared photon can penetrate
 the whole ISM of dIrrs. Considering the physical process which
 timescale is the longest as main process in regulating global star
 formation, we can show this lack of \hmol leads the quiescent phase of
 star formation. Hence, we can say that the stellar radiation which
 destroys \hmin and leads low \hmol abundance should be properly treated
 in studying SFHs of dIrrs.

 }

\titlerunning{The self regulating star formation of gas rich dwarfs}
\authorrunning{MARK \& H. Kamaya}

\maketitle

 \section{INTRODUCTION}

 Dwarf galaxies are the galaxies with lower luminosity, lower mass and
 smaller size, and they are much more abundant in number than normal
 galaxies on the Hubble sequence like our Milky Way \citep[see][for a
 recent review of local group dwarf galaxies]{ko00}. These natures imply
 that they are basic building blocks of the normal galaxies in
 hierarchical structure formation scenarios of the cold dark matter
 universe \citep{clbf00, cv02, mo03}. In addition, the dwarf galaxies
 have lower abundances of heavy elements than those of the normal
 galaxies; therefore, they are still in an early stage of chemical
 evolution and/or are analogies to primordial galaxies
 \citep[e.g.,][]{it99}.

 Dwarf galaxies are classified by their morphological appearances into
 some basic types. One of the morphological types is dwarf irregular
 galaxies (dIrrs), which characteristics are irregular and amorphous
 appearances at optical wavelengths. Interestingly, unlike other
 morphological types of dwarf galaxies, dIrrs still have plentiful
 interstellar medium (ISM) and in general show ongoing star formation
 and HII region \citep{ma98, ko00}. The star formation histories (SFHs)
 of dIrrs are considered to be episodic, that is, quiescent phase
 follows a starburst phase \citep{greggio93, tolstoy98, grebel99}, and
 their star formation rates (SFRs) are more widely distributed ($\sim
 10^{-4}-10^{-2}\dsfr$) than that of the normal galaxies
 \citep{hunter97}. In this issue, we particularly focus on the dIrrs
 which are considered to be in their quiescent phase of star formation
 (i.e. $\sfr \sim 10^{-4}\dsfr$).

 However, despite their closeness, the regulation of star formation even
 in the Local Group dIrrs is not yet well understood. Actually, it is
 very surprising that they have been forming stars recently because
 their gravitational potential are shallow \citep[][for a review]{sb97}
 and comparable to the energies of some supernova explosions; therefore,
 they should have little ISM after some supernovae (several million
 years from the epoch of dIrrs formed), and thus, the recent and ongoing
 star formation would not occur. Why do they still have plentiful gas?
 Why do they show ongoing star formation? Both the questions are still
 controversial. Even if these questions are solved, another puzzle
 remains; why do they form stars at such low rates in spite of having
 plentiful ISM as material of stars? This paper attempts to solve this
 puzzle by investigating the non-chemical equilibrium molecule formation
 history of dIrrs on some assumptions.

 For the star formation processes in low-metallicity environment,
 hydrogen molecules ($\mathrm{H_2}$) can be important coolants at the
 gas temperature $T\la 10^4\dkel$ \citep{pd68}; more abundant \hmol
 leads gas to cool faster, and therefore, would result in more active
 star formation. Although \hmol forms primarily on the surfaces of dust
 grains in local molecular clouds, \hmol can also form in gas-phase
 through the reactions, which intermediaries are negatively charged
 hydrogen ($\mathrm{H^-}$),
 \begin{eqnarray}
  \hneut +\mathrm{e^-} &\rightarrow & \mathrm{H^-}+\gamma , 
   {\label{hm1}}\\
  \mathrm{H^-}+\hneut &\rightarrow & \mathrm{H_2}+\mathrm{e^-},
   {\label{hm2}}
 \end{eqnarray}
 or ionized hydrogen molecule ($\mathrm{H_2^+}$),
 \begin{eqnarray}
  \hneut +\mathrm{H^+} &\rightarrow & \mathrm{H_2^+} +\gamma ,
   {\label{h2p1}}\\
  \mathrm{H_2^+}+\hneut &\rightarrow & \mathrm{H_2} + \mathrm{H^+}. 
   {\label{h2p2}}
 \end{eqnarray}
 These gas-phase reactions are active $T\sim 10^4\dkel$ and the
 reactions ({\ref{hm1}}) and ({\ref{hm2}}) can more efficiently form
 \hmol than the ({\ref{h2p1}}) and ({\ref{h2p2}}) reactions
 \citep[e.g.,][]{abel97}. On the other hand, \hmol is dissociated by
 collisions with other chemical species and the absorption of
 ultraviolet (UV) photons called the Lyman-Werner band (LW) photons
 ($11.26\ \mathrm{eV} \le h\nu \le 13.6\ \mathrm{eV}$), which are mainly
 radiated by early type stars \citep{fsd66}.
  
 Many astrophysicists have built the theoretical models concentrated on
 the dissociation of \hmol by the LW photons because they have
 investigated the second epoch of star formation after the first stars
 have been formed in metal-free molecular cloud \citep[e.g.,][]{silk77,
 nt00}. These models result in the low abundance of \hmol in relatively
 large region (several hundred parsec around the source star of the LW
 photons), and thus, negative-feedback on further star formation in the
 host cloud of the star. However, in dIrrs, the region affected by LW
 photons would be smaller than previously predicted because dIrrs are
 gas rich system; the mean free paths of LW photons are on the order of
 $1\ \mathrm{pc}$ ($\ll$ the typical size of dIrrs) with typical
 parameters of primordial clouds \citep{hrl96}. Furthermore, early type
 stars as the sources of LW photons are rare in dIrrs. Therefore, the
 negative-feedback on further star formation brought about the \hmol
 dissociation by the LW photons may be negligible in dIrrs.

 By the way, if \hmol is formed by gas-phase reactions
 (equations~\ref{hm1}-{\ref{h2p2}}), the formative abundance of
 $\mathrm{H_2}$, $y_\mathrm{H_2}$, largely depends on the \hmin
 abundance, $y_\mathrm{H^-}$, because the \hmol formation process
 through \hmin is more effective than the one through $\mathrm{H_2^+}$
 (in \S2, we discuss that gas-phase \hmol formation may dominate
 grain-catalyzed \hmol formation in dIrrs). We expect that \yhmol in
 dIrrs is more effectively influenced by the destruction of \hmin than
 by the direct dissociation of \hmol due to the LW photons. This is
 because that the typical mean free path of the \hmin destroying
 infrared (IR) photons (its wavelength $\lambda \la 1.64\ \mathrm{\mu
 m}$), $l_\mathrm{IR}$, would be sufficiently longer than that of the LW
 photons; $l_\mathrm{IR}\sim 100\ \mathrm{kpc}$ ($\gg$ the typical size
 of dIrrs) with typical parameters of primordial clouds. Therefore, the
 IR photons can penetrate and affect the whole ISM. Furthermore, the IR
 photons are also radiated by low-mass stars as well as early type
 stars. Thus, all stars born in the previous starburst phase can be
 considered as the sources of the \hmin destroying photons. If \hmin is
 seriously destroyed by the IR photons of the interstellar radiation
 field (ISRF) of dIrrs, the star formation activity may become more
 inactive at almost the whole ISM.

 With this expectation, we construct the theoretical model to
 investigate $y_\mathrm{H_2}$. In \S2, we compare the rates of gas-phase
 and grain-catalyzed \hmol formation, and show gas-phase \hmol formation
 may dominate in dIrrs. \S 3 describes our model for the star formation
 process in dIrrs and \S 4 presents our results. Finally, we summarize
 our study and briefly discuss about the next starburst phase in \S 5.

 \section{GAS-PHASE AND GRAIN-CATALYZED \hmol FORMATION}

 \hmol formation on dust-grain surfaces completely dominates gas-phase
 one in the local ISM. However, if the dust-to-gas ratio, $D$, is lower
 than a critical value, $D_\mathrm{cr}$, gas-phase \hmol formation
 dominates grain-catalyzed formation \citep{kh01, glover03}. Despite the
 \hmol formation rate on grain surfaces is still uncertain,
 \citet{glover03} calculates \dcr in some astrophysical situations by
 tentatively adopting the rate of \citet{hm79}. As noted in
 \citet{glover03}, the temperature dependence of \dcr is very strong; at
 low temperatures ($T\la \mathrm{a\ few}\times 10^2\dkel$),
 grain-catalyzed \hmol formation is relatively efficient and \dcr is
 very small ($D_\mathrm{cr}\la 10^{-3}D_\mathrm{MW}$, where \dmw is the
 dust-to-gas ratio in the local ISM), while at high temperatures,
 gas-phase \hmol formation is efficient and thus, \dcr is large. In
 addition, \dcr also depends on the ionisation degree and the density of
 ISM; both high-ionisation and high-density lead the high efficiency of
 gas-phase \hmol formation and large $D_\mathrm{cr}$.

\begin{figure}
 \resizebox{\hsize}{!}{\includegraphics{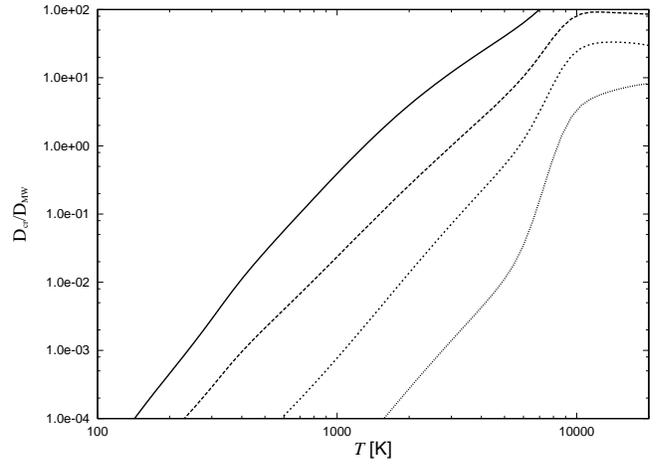}}
 \caption{\dcr as a function of the gas temperature $T$ under some ISRF
 intensities are presented. The thick, long-dashed, short-dashed and
 dotted lines represent \dcr for the ISRF intensity ratios $\varepsilon$
 of 0.010, 0.10, 1.0, and 10.0, respectively. The ionisation degree and
 the density are self-consistently calculated.}
 \label{fig1}
\end{figure}

 By the way, we focus on the process that the ISM of dIrrs once is
 heated up by stellar-radiation, supernovae and/or other processes, and
 then cools down to the temperature at which star formation is effective
 under ISRF. Thus, re-calculating \dcr (eq. (35) of \citet{glover03}) by
 using the physical parameters in which we interested, we have
 $D_\mathrm{cr} \ga 10^{-2}\dmw$ at $T\ga 5000\dkel$ as shown in the
 dotted line of figure~\ref{fig1}. This critical value is on the same
 order of magnitude with the typical value for dIrrs
 $D_\mathrm{dIrrs}\sim 1.3\times 10^{-2}\dmw$ \citep{lf98}.

 At $T\la 5000\dkel$, the ISM cools off at a stroke. This is because
 $D_\mathrm{cr}$ becomes smaller than $D_\mathrm{dIrrs}$, and thus,
 \hmol formation on dust-grain surfaces may dominate; therefore, \yhmol
 rapidly increases. Furthermore, if the metallicity of dIrrs is higher
 than a critical value $\sim 10^{-2}Z_{\sun}$ \citep{nt00} (the same
 order with the metallicity of the most metal-deficient galaxy, I Zw
 18), metal cooling dominates \hmol cooling at this temperature
 range. However, we are interested in the temperature at which the
 cooling timescale is the longest because global star formation
 timescale is decided by the slowest physical process. Thus, the
 interesting temperature is found to be $\sim 5000\dkel$, at which
 gas-phase \hmol formation dominates. Therefore, it is meaningful to
 study the SFHs of dIrrs on the assumption that gas-phase \hmol
 formation is important.

 \section{MODEL AND CALCULATION}

 We calculate the time-dependent, non-chemical equilibrium, ionisation
 and molecule formation history of a gas system in dIrrs composed of
 hydrogen and helium. Neglecting the dynamics of the gas system in ISM,
 we concentrate on the time evolution of the density and temperature in
 some regions in dIrrs. We consider the following 9 chemical species;
 $\mathrm{H,\ H^+,\ H^-,\ H_2,\ H_2^+,\ He,\ He^+,\ He^{++},\ e^-}$. In
 dIrrs, the helium abundance is still maintained almost primordial value
 because dIrrs are still in an early stage of chemical evolution;
 therefore, it is assumed to be primordial value (24\% of hydrogen by
 mass fraction) in our calculation \citep{skillman93, izotov99}. 

 At first, the ISRF in our model, $J_{\nu}$, is assumed to have same
 energy dependence with that of the solar vicinity, $J_{\nu}^{\sun}$
 \citep[see][]{mmp83}, and its intensity is normalised by that of the
 solar vicinity; $J_{\nu} = \varepsilon \cdot J_{\nu}^{\sun}$. In order
 to investigate the dependence of \yhmol on the ISRF, we change the
 ratio $\varepsilon$ from 0.01 to 100 with the step of one order of
 magnitude. At next, decomposing the ISRF of solar vicinity into four
 parts, that is, early type stars, two types of disc stars ($\teff =
 7500\dkel,\ 4000\dkel$) and red giant stars ($\teff = 3000\dkel$) as in
 \citet{mmp83}, we investigate which degree each component affects
 $y_\mathrm{H_2}$. Initial conditions of all calculations are
 $T_\mathrm{ini} = 1.0\times 10^6\dkel,\ n_\mathrm{ini}=3.0\times
 10^{-3}\ \mathrm{cm^{-3}}$, and the initial abundances of each chemical
 species are that in chemical equilibrium at $T_\mathrm{ini}$.

 \section{Results} 

\begin{figure}
 \resizebox{\hsize}{!}{\includegraphics{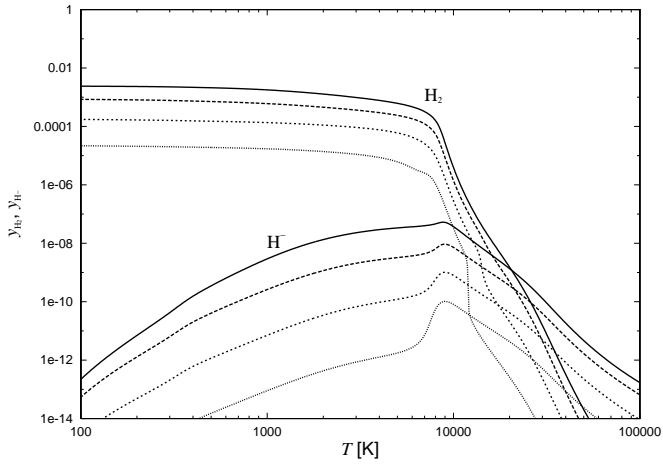}}
 \caption{\yhmol and \yhmin as a function of $T$ under some ISRF
 intensities are presented. The four lines increased monotonically as
 the temperature decreases are \yhmol (labelled as "$\mathrm{H_2}$") and
 other four lines are \yhmin (also labelled as "$\mathrm{H^-}$"). The
 thick, long-dashed, short-dashed and dotted lines represent the
 formative abundances for the ISRF intensity ratios $\varepsilon$ of
 0.010, 0.10, 1.0, and 10.0, respectively.}
 \label{fig2}
\end{figure}

 Figure~\ref{fig2} shows \yhmol and \yhmin under some ISRF
 intensities. As the ISRF becomes more intense, these formative
 abundances monotonically decrease at a fixed temperature. \yhmol at
 $T=5000\dkel$, which is tightly connected with the SFRs of dIrrs in our
 model (see equation~\ref{eq-sfr2}), reaches $\sim 10^{-5}$ for the ISRF
 ten times more intense than that of the solar vicinity. This low \hmol
 abundance means that the ISM cannot cool down and that star formation
 is delayed, and thus, this can explain the observed small SFRs of dIrrs
 in their quiescent phase.

\begin{figure}
 \resizebox{\hsize}{!}{\includegraphics{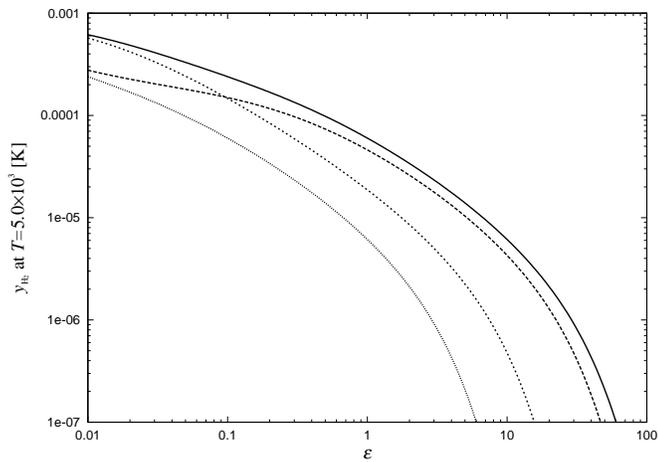}}
 \caption{\yhmol at $T = 5000\dkel$ as a function of the normalized
 ISRF intensity, $\varepsilon$, and its dependence on the spectral
 energy distribution (SED) of ISRF are shown. The thick line represents
 \yhmol for the SED of the solar vicinity ISRF. The long-dashed line is
 also the one, while the ISRF has increment of all components except for
 early type stars by an order of magnitude. The short-dashed is that for
 the ISRF with increment of only the component of early type stars by an
 order of magnitude. The dotted line is for the all components to be
 increased by an order of magnitude.}
 \label{fig3}
\end{figure}

 The dependences of \yhmol at $5000\dkel$ on the ISRF intensity are
 drawn in figure~\ref{fig3}. The radiation from both disc stars ($\teff
 = 7500\dkel,\ 4000\dkel$) and giant stars ($\teff = 3000\dkel$) is as
 important as that from early type stars as the LW band photon sources
 in order to decrease $y_\mathrm{H_2}$. The ISRF intensity, which is
 slightly stronger than that of solar vicinity, may be easily attainable
 because the ISRF intensity of the solar vicinity is still much weaker
 than that of the Magellanic Clouds \citep{bb01, contursi00}.

 Here, we simply model the SFR, using Jeans mass, \jmass, Jeans radius,
 \jrad, cooling time, \tcool, and the mass conversion fraction from
 clouds to stars (star formation efficiency; \sfe);
 \begin{equation}
  \sfr =\frac{\sfe \cdot \jmass}{\tcool}/{\jrad}^2. 
   {\label{eq-sfr1}} 
 \end{equation}
 This equation means that a cloud with mass $M_\mathrm{J}$ and size
 $r_\mathrm{J}$ will cool down in the timescale $t_\mathrm{cool}$ and
 some portions of the cloud which mass fraction are represented by SFE
 will convert into stars. Then, we apply the model to our simulation,
 and get
 \begin{equation}
  \sfr \sim 1.0\times 10^{-4}\cdot \left(\frac{\sfe}{10^{-3}} \right) 
   \cdot \left(\frac{\yhmol}{10^{-5}} \right) \dsfr,
   {\label{eq-sfr2}}
 \end{equation}
 where \yhmol is the abundance of \hmol at $5000\dkel$, at which the
 cooling timescale is the longest. We can predict that the SFE of dIrrs
 is low (the order of $10^{-3}$), and ISRF of dIrrs is stronger than
 that of the solar vicinity in order to understand why the SFRs of dIrrs
 are such low.

 \section{SUMMARY AND DISCUSSION}
 
\begin{figure}
 \resizebox{\hsize}{!}{\includegraphics{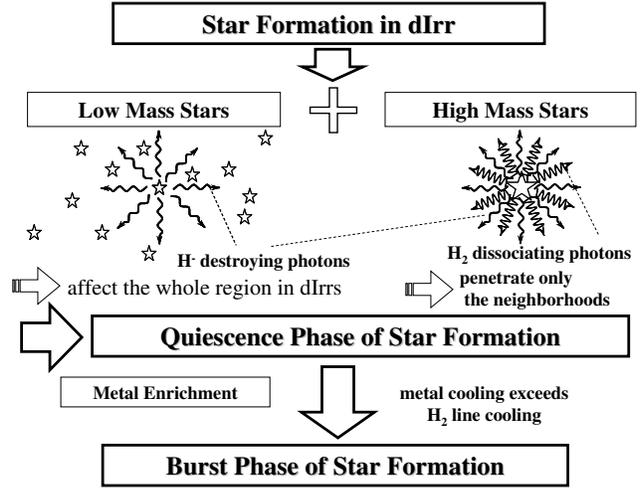}}
 \caption{The schematic explanation of our scenario for SFHs of dIrrs.}
 \label{fig4}
\end{figure}

 We can conclude that the low mass stars, which are considered to be
 \hmin destroying photon sources, are as important as high mass OB stars
 as the sources of LW band photon in understanding the episodic SFHs of
 dIrrs. That is, the radiation from many low mass stars burn in previous
 burst epoch of star formation destroys $\mathrm{H^-}$, which is the
 intermediary of gas-phase \hmol formation process, and suppresses
 further \hmol formation with \hmol dissociation, and keep the SFHs in
 quiescent phase. The schematic explanation of our conclusion is
 represented by figure~\ref{fig4}. Incidentally, we can say that low
 mass stars are more important than OB stars because the typical life
 time of OB stars is considerably short ($\sim 10^6$ yr). Furthermore,
 low mass stars may be much more abundant in low surface brightness
 galaxies than ordinary expected \citep{lee04}; that is,
 ``bottom-heavy'' initial mass function (IMF) is suggested there. If
 this IMF stands up in dIrrs, the contribution of the low mass stars to
 decrease \yhmol may become more and more important.


\begin{acknowledgements}
We are grateful to the referee for his/her encouraging comments.
This work is supported by the Grant-in-Aid from the Ministry of 
Education, Culture, Sports, Science and Technology (MEXT) of Japan
(16740110) and the Grant-in-Aid for the 21st Century COE
"Center for Diversity and Universality in Physics" from MEXT of Japan.
\end{acknowledgements}


\bibliographystyle{aa}
\bibliography{}
\end{document}